\documentclass[12pt]{iopart}
\usepackage{iopams}  
\usepackage{graphicx}

\begin{document}

\title[]{Adsorption of water on the pristine and defective semiconducting 2D CrPX$_3$ monolayers (X: S, Se) }

\author{Sifan Xu$^1$, Zhicheng Wu$^1$, Yuriy Dedkov$^{1,2,*}$, and \newline Elena Voloshina$^{1,2,*}$}

\address{$^1$Department of Physics, Shanghai University, 99 Shangda Road, 200444 Shanghai, China}
\address{$^2$Institut f\"ur Chemie und Biochemie, Freie Universit\"at Berlin, Arnimallee 22, 14195 Berlin, Germany}


\vspace{10pt}
\vspace{10pt}


\begin{abstract}

The effect of vacancy and water adsorption on the electronic structure of semiconducting 2D trichalcogenide material CrPX$_3$ (X: S, Se) is studied using state-of-the-art density functional theory (DFT) approach. It is found that chalcogen vacancies play a minor role on the electronic structure of CrPX$_3$ in the vicinity of the Fermi level leading to the slightly reduced band gap for these materials, however, inducing strongly localised defect states which are placed in the energy gap formed by the valence band states. Our DFT calculations show that the interaction of water molecules with CrPX$_3$, pristine and defective, can be described as physisorption and the adsorption energy for H$_2$O is insensitive to the difference between pristine and chalcogen-defective surface of trichalcogenide material. These results are the first steps for the theoretical description of the ambient molecules interaction with 2D semiconducting CrPX$_3$ material, that is important for its future experimental studies and possible applications.

\end{abstract}

\section{Introduction}

The class of the pure 2D and quasi-2D materials is rapidly developing~\cite{Zeng:2018bz,Akinwande:2019cn,Dedkov:2020da,Andrei:2020fl} regularly bringing new exciting phenomena, like observation of the 2D magnetic order~\cite{Lee:2016ga,Wang:2016ez,Gong:2017jf}, superconductivity~\cite{Cao:2018ff,Cao:2018kn}, and exciting optical properties~\cite{Kang:2020fx,Ma:2021ho}. Among these materials, quasi-2D transition metal trichalcogenides (MPX$_3$, M: transition metal, X: S, Se)  recently attracted increased attention because of their possible applications in many areas. It was proposed that they can be used as low-dimensional spin-polarised conductors which spin character can be tuned by the applied bias~\cite{Li:2014de}, in photocatalysis for the efficient water splitting and hydrogen production~\cite{Cheng:2018co,Cheng:2020ki,Du:2018dh}, as efficient materials for the Li storage~\cite{Fan:2019gh},  and, for example, the tuning of the vacancy state can lead to the low-dimensional ferromagnetic state~\cite{Bai:2020ct}.

From the crystallographic point of view, these materials are similar to, e.\,g., MoS$_2$, where one third of the metal atoms is substituted by P--P dimers and in this case the metal layer in MPX$_3$ is encapsulated by both chalcogens and phosphorus atoms (Fig.~\ref{fig:structure}). Diversity in transition metals and two different chalcogens can lead to a wide variation in the electronic and magnetic properties of these materials. Also, the simultaneous presence of sulphur and phosphorus in these compounds can cause a synergetic effect on the electronic structure of the central metal atoms. Moreover, due to the layered structure of MPX$_3$, they can be easily prepared as two-dimensional nanostructures having a large surface area with large number of active sites. Surprisingly, the electronic structure of these materials is mainly addressed only from the theory side and the experimental studies are very rear and sporadically appear in the literature~\cite{Yan:2021cn}, moreover not always correctly interpreting the obtained experimental data~\cite{Dedkov:2020ca,Gusmao:2017kz}. Also, the influence of vacancies as well as ambient adsorbates on the magnetic and electronic properties of MPX$_3$ are not studied at all.

Here, we present state-of-the-art density functional theory (DFT) studies of Cr-based phosphorus trichalcogenides (CrPS$_3$ and CrPSe$_3$). These materials were recently synthesised and studied with respect to their electrochemical sensing and energy applications~\cite{Gusmao:2017kz}. In our work we address many aspects of these materials, like stability, chalcogen defects formation and adsorption of water molecules on pristine and defective 2D CrPX$_3$ layer. We found that in all cases, the electronic structure around the top of the valence band of CrPX$_3$ remains insensitive and main modifications are connected with the valence states redistributions at large binding energies. These results demonstrate that CrPX$_3$ are robust to the external factors, like chalcogen defects formation and ambient molecules adsorption, that can be utilised in future applications of these materials as, e.\,g., protective coatings or inert barriers for molecules.   

\section{Computational details}

Spin-polarised DFT calculations based on plane-wave basis sets of $500$\,eV cutoff energy were performed with the Vienna \textit{ab initio} simulation package (VASP).~\cite{Kresse:1996kg,Kresse:1994cp,Kresse:1993hw} The Perdew-Burke-Ernzerhof (PBE) exchange-correlation functional~\cite{Perdew:1996ky} was employed. The electron-ion interaction was described within the projector augmented wave (PAW) method~\cite{Blochl:1994fq} with Cr ($3p$, $3d$, $4s$), P ($3s$, $3p$), S ($3s$, $3p$) and Se ($4s$, $4p$) states treated as valence states. The Brillouin-zone integration was performed on $\Gamma$-centred symmetry reduced Monkhorst-Pack meshes using a Gaussian smearing with $\sigma = 0.05$\,eV, except for the calculation of total energies. For these calculations, the tetrahedron method with Bl\"ochl corrections~\cite{Blochl:1994ip} was employed. The $12\times12\times4$ and $24\times24\times1$ $k$-meshes were used for the studies of bulk and monolayer CrPX$_3$, respectively, and the $12\times12\times1$ $k$-mesh was used for the ($2\times2$) supercells consisting of $4$-fold unit monolayers.  The DFT+$\,U$ scheme~\cite{Anisimov:1997ep,Dudarev:1998dq} was adopted for the treatment of Cr $3d$ orbitals, with the parameter $U_\mathrm{eff}=U-J$ equal to $4$\,eV. Dispersion interactions were considered adding a $1/r^6$ atom-atom term as parameterised by Grimme (``D2'' parameterisation)~\cite{Grimme:2006fc}. This approach yields structural parameters, which are in good agreement with available experimental data~\cite{Gusmao:2017kz}. 

When modelling CrPX$_3$ monolayers, the lattice constant in the lateral plane was set according to the optimised lattice constant of bulk CrPX$_3$. A vacuum gap was set to approximately $26$\,\AA. During structure optimisation, the convergence criteria for energy and force were set equal to $10^{-5}$\,eV and $10^{-2}$\,eV/\AA, respectively. 

The cleavage energy was calculated as 
\begin{equation}\nonumber
E_\mathrm{cl} =\left(E_{d_0\rightarrow\infty} - E_0\right)/A,
\label{equ:	cleav}
\end{equation}
where $d_0$ is the van der Waals gap of bulk crystal and $A$ is the in-plane area.

To extract the exchange interaction parameters between Cr ions spins, the Heisenberg Hamiltonian was considered
\begin{equation}\nonumber
H=\sum_{\langle i,j \rangle}J_1\vec{S}_i \cdot \vec{S}_j+\sum_{\langle \langle i,j \rangle \rangle}J_2\vec{S}_i \cdot \vec{S}_j+\sum_{\langle \langle \langle i,j \rangle \rangle \rangle}J_3\vec{S}_i \cdot \vec{S}_j,
\end{equation}
where $\vec{S}_i$ is the net spin magnetic moment of the Cr ions at site $i$, three different distance magnetic coupling parameters were estimated, considering one central Cr ions interacted with three nearest neighbouring (NN, $J_1$), six next-nearest neighbouring (2NN, $J_2$), and three third-nearest neighbouring (3NN, $J_3$) Cr ions, respectively. Here, the long-range magnetic exchange parameters ($J$) can be obtained as~\cite{Sivadas:2015gq}
\begin{eqnarray}\nonumber
\label{equ:J}
& J_1=\frac{(E_\mathrm{sAFM}-E_\mathrm{zAFM})+(E_\mathrm{nAFM}-E_\mathrm{FM})}{16S^2}\,,\\ 
&J_2=\frac{(E_\mathrm{zAFM}+E_\mathrm{sAFM})-(E_\mathrm{nAFM}+E_\mathrm{FM})}{32S^2}\,, \\ \nonumber
&J_3=\frac{3\,(E_\mathrm{zAFM}-E_\mathrm{sAFM})+ (E_\mathrm{nAFM}+E_\mathrm{FM})}{48S^2}\,,
 \end{eqnarray}
where $S$ is the calculated magnetic moment of the Cr ion and $E_\mathrm{FM}$, $E_\mathrm{nAFM}$, $E_\mathrm{zAFM}$, $E_\mathrm{sAFM}$ are the total energies in ferromagnetic, N\'eel antiferromagnetic, zigzag antiferromagnetic, and stripy antiferromagnetic configurations, respectively.

To estimate $T_\mathrm{N}$ temperature, Monte Carlo simulations were performed within the Metropolis algorithm with periodic boundary conditions~\cite{Metropolis:1953in}. The three exchange parameters $J_1$, $J_2$ and $J_3$ were used in a $64\times 64$ superlattice containing a large enough amount of magnetic sites to accurately evaluate the value. Upon the heat capacity $C_v(T) = (\langle E^2\rangle - \langle E\rangle^2)/k_B T^2$ reaching the equilibrium state at a given temperature, the $T_\mathrm{N}$ value can be extracted from the peak of the specific heat profile.

The electrically neutral vacancies were created by removing one or two X atoms from the ($2\times 2$) supercells. Thereby, the distance between repeated vacancies in the nearest-neighbour cells is larger than $10$\,\AA. The defect formation energy is defined as follows
\begin{equation}\nonumber
\Delta E_\mathrm{def} = \frac{1}{n}\left[E(\mathrm{CrPX}_{3-n})+ n\,\mu_\mathrm{X} - E(\mathrm{CrPX}_3)   \right]\,,
\label{equ:defects}
\end{equation}
where $n$ is a number of defects, $E(\mathrm{CrPX}_{3-n})$ and $E(\mathrm{CrPX}_3)$ are the energies of the 2D CrPX$_3$ with and without vacancy, respectively, $\mu_\mathrm{X}$ is the chemical potential of X atom ($\mu_\mathrm{S}=-4.1279$\,eV and $\mu_\mathrm{Se}=-3.4895$\,eV). 

To study the adsorption of a single molecule, a ($2\times2$) supercell was used with one water molecule added. Adsorption energies were calculated as
\begin{equation}\nonumber
\Delta E_\mathrm{ads} = E(\mathrm{A}/\mathrm{CrPX}_3) - \left[E(\mathrm{A}) + E(\mathrm{CrPX}_3)\right]\,,
\label{equ:adsorption}
\end{equation}
where $E(\mathrm{CrPX}_3)$ and $E(\mathrm{A})$ are he energies of the isolated 2D CrPX$_3$ and an adsorbate, and $E(\mathrm{A}/\mathrm{CrPX}_3)$ is the energy of their interacting assembly.

\section{Results and Discussion}

\subsection{3D CrPX$_3$ (X = S, Se)}

\begin{figure}[h]
\begin{center}
\includegraphics[width=1.0\textwidth]{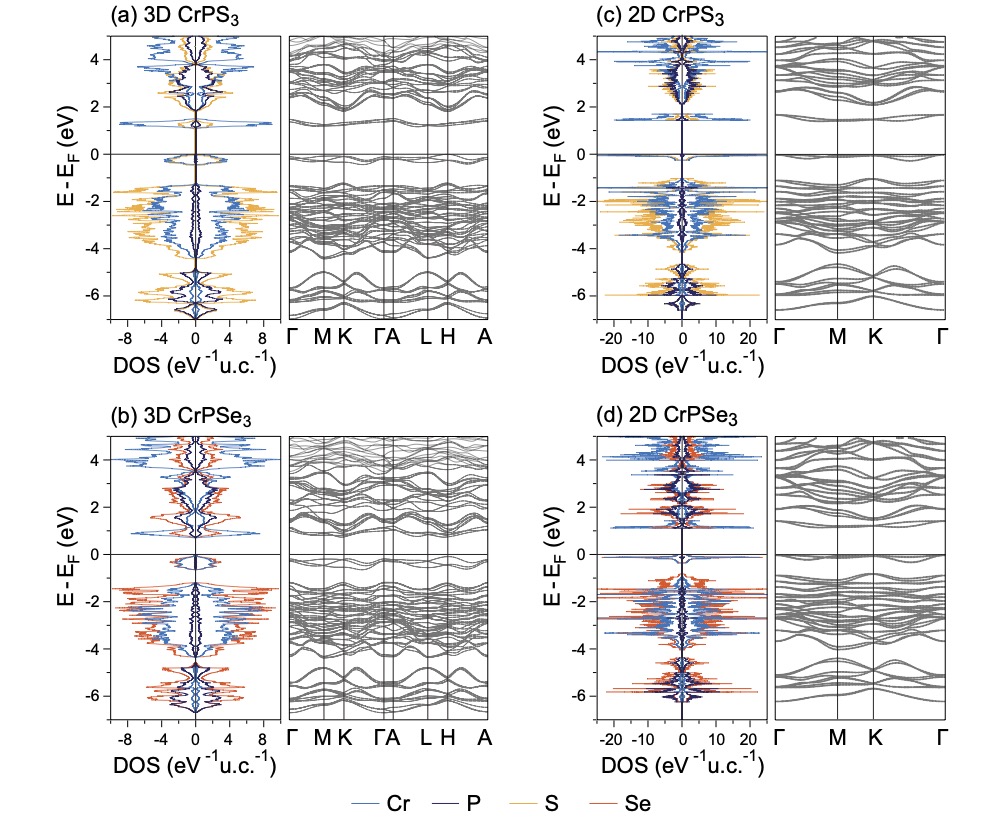}
\end{center}
\caption{Band structure and site-projected density of states for (a) 3D CrPS$_3$ bulk, (b) 3D CrPSe$_3$ bulk, (c) 2D CrPS$_3$ monolayer, and (d) 2D CrPSe$_3$ monolayer. }
\label{fig:dos3D}
\end{figure}

3D bulk MPX$_3$ crystals usually adopt either an $AAA$ stacking in $C2/m$ space group space or $ABC$ stacking in $R\bar3$ space group. Both of them can be represented in hexagonal unit cells (Fig.~\ref{fig:structure}\,a). Then every unit cell contains three MPX$_3$ single layers which have $D_{3d}$ symmetry (see Fig.~\ref{fig:structure}\,b,c), which are stacked in a different way. Each single-layer are held together by van der Waals forces. The lattice parameters of hexagonal 3D CrPX$_3$ were fully relaxed in nonmagnetic (NM), ferromagnetic (FM) and N\'eel antiferromagnetic (nAFM) states  and they are listed in Table~\ref{tab:3D} with the respective total energies. Thus, both CrPX$_3$  (X = S, Se) prefer the $C2/m$ symmetry, that is in agreement with available experimental data~\cite{Brec:1980ts}. For both CrPS$_3$ and CrPSe$_3$, the energetically most favourable structure corresponds to the nAFM configuration (Tab.~\ref{tab:3D}). The band structure and density of states (DOS) calculated for the ground-state structures of 3D CrPS$_3$ and CrPSe$_3$ are shown in Figure~\ref{fig:dos3D}\,(a) and (b), respectively. From these results one can see, that the both systems under study are indirect band gaps semiconductors. The band gaps obtained by the PBE$+U$ method are $1.11$\,eV for bulk CrPS$_3$ and $0.70$\,eV for bulk CrPSe$_3$. The upper valence bands (VBs) are mainly formed by S/Se $p$ and equal to that by the Cr $3d$ orbitals; the lower conduction bands (CBs) are composed of Cr $3d$ states with somewhat contribution from S/Se $p$. The partial contributions from P $p$ orbitals are very small in VBs as well as CBs. These results are very different from the respective data known for MnPX$_3$~\cite{Yang:2020ex} and NiPX$_3$~\cite{Yan:2021cn,Wu:2021ac}, where the contribution of metal $3d$ states in VBs was insignificant.

\begin{figure}[h]
\begin{center}
\includegraphics[width=1.0\textwidth]{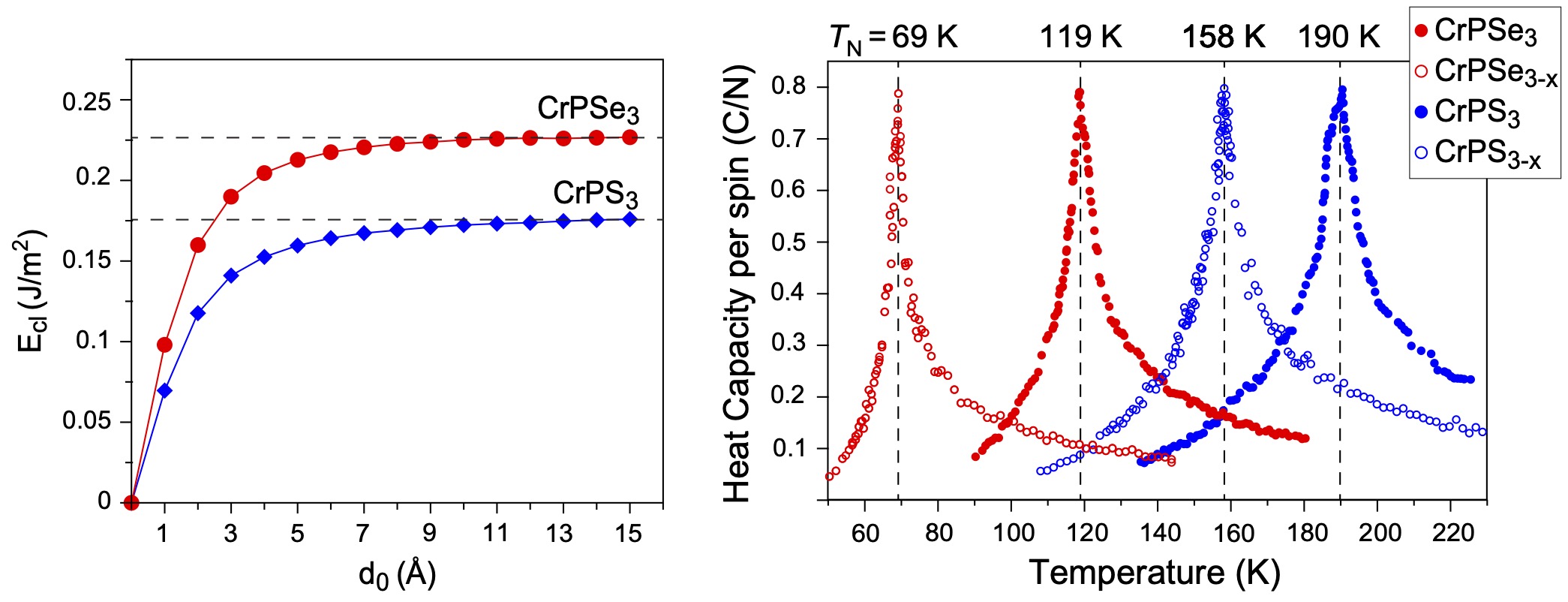}
\end{center}
\caption{(Left panel) Cleavage energy of 3D CrPX$_3$ ($E_\mathrm{cl}$) as a function of the van der Waals gap ($d_0$). (Right panel) Specific heat capacity per spin with respect to temperature for  2D CrPX$_3$ and CrPX$_{3-x}$. }
\label{fig:cleav+MC}
\end{figure}

It is expected that individual CrPX$_3$ layers can be isolated by mechanical exfoliation. In this regard, the cleavage energy ($E_\mathrm{cl}$) of a layered material is an essential property that need to be considered. The calculated values, $E_\mathrm{cl}(\mathrm{CrPS}_3) = 0.18$\,J\,m$^{-2}$ and $E_\mathrm{cl}(\mathrm{CrPSe}_3) = 0.23$\,J\,m$^{-2}$ (Fig.~\ref{fig:cleav+MC}, left panel), are smaller than that of a bulk graphite ($0.36$\,J\,m$^{-2}$)~\cite{Zacharia:2004go}, which is used as an indicator for the feasibility of exfoliation of materials in experiments. The obtained values are in the same range as recently published for MnPX$_3$ and NiPX$_3$~\cite{Yang:2020ex,Wu:2021ac}.

\subsection{2D CrPX$_3$ (X = S, Se) monolayers}

\begin{figure}[h]
\begin{center}
\includegraphics[width=1.0\textwidth]{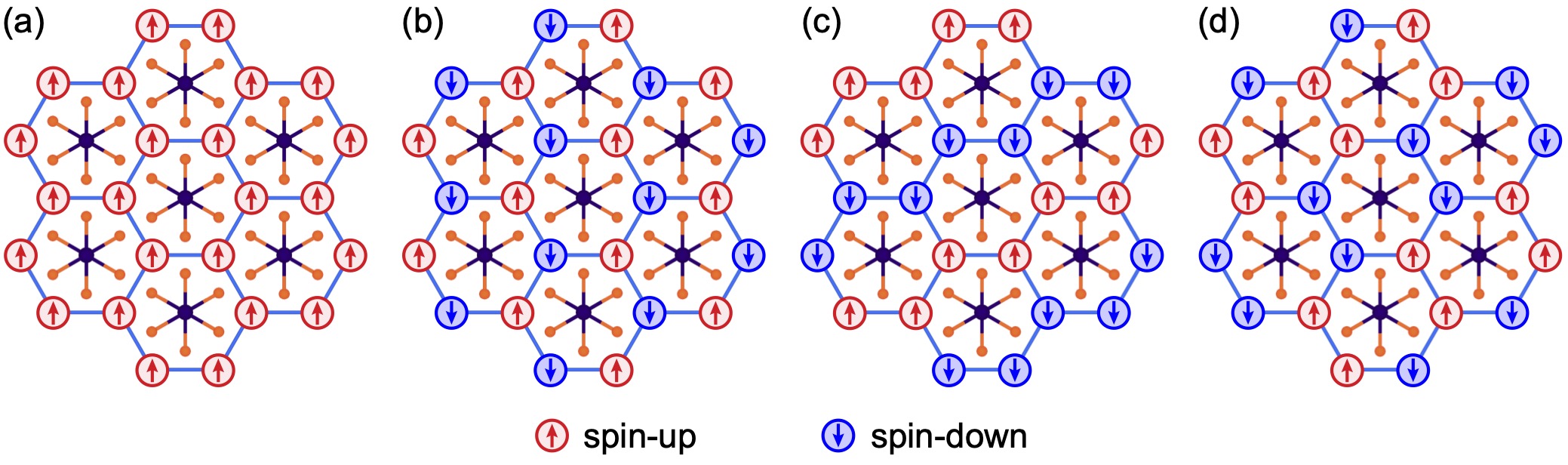}
\end{center}
\caption{Four different magnetic configurations of 2D CrPX$_3$: (a) ferromagnetic (FM), (b) N\'eel antiferromagnetic (nAFM), (c) zigzag antiferromagnetic (zAFM), and (d) stripy antiferromagnetic (sAFM).}
\label{fig:magnetic}
\end{figure}

Four possible magnetic configurations were investigated to evaluate the ground state of 2D CrPX$_3$ monolayers: The mentioned above FM (Fig.~\ref{fig:magnetic}\,a), and nAFM (Fig.~\ref{fig:magnetic}\,b) were supplemented by a zig-zag AFM (zAFM, Fig.~\ref{fig:magnetic}\,c) and a stripy AFM (sAFM, Fig.~\ref{fig:magnetic}\,d). The obtained results are listed in Table~\ref{tab:2D}. It can be seen that the lowest energy configurations for the 2D CrPX$_3$ (X = S, Se) monolayers are the N\'eel AFM states. 

The 2D CrPS$_3$ and CrPSe$_3$ in their ground states keep an indirect semiconductor behaviour (Fig.~\ref{fig:dos3D}c,d). Due to the quantum confinement effects, the band gaps are slightly larger than the values calculated for the bulk phases and are $1.52$\,eV and $1.11$\,eV for the 2D CrPS$_3$ and CrPSe$_3$, respectively. As for the bulk materials, the bands in the vicinity of Fermi energy are mostly composed of Cr-$d$ and S/Se-$p$ orbitals.  

The exchange-coupling parameters ($J_1$, $J_2$, $J_3$) describing the magnetic interactions between Cr$^{2+}$ ions were calculated according to eqn.~(\ref{equ:J}) and are listed in Tab.~\ref{tab:2D}. According to the Goodenough-Kanamori-Anderson (GKA) rules~\cite{Goodenough:1955th,Kanamori:1960ki},  the origin of the nAFM ground state can be attributed to the competition between the NN AFM direct Cr--Cr ($d-d$) exchange interactions and the indirect Cr--Se$\,\cdots\,$Se--Cr ($p-d$) superexchange interactions~\cite{Yang:2020ko}. The calculated $J$ parameters were used in the Monte-Carlo simulations of N\'eel temperatures of CrPX$_3$ monolayers. The estimated $T_\mathrm{N}$ is $190$\,K for 2D CrPS$_3$ and $119$\,K for 2D CrPSe$_3$ (Fig.~\ref{fig:cleav+MC}, right panel). These results are in very good agreement with the experimental result available for CrPSe$_3$, that is: $T_\mathrm{N}=136$\,K~\cite{Gusmao:2017kz}.

\subsection{Chalcogen defects}

\begin{figure}[h]
\begin{center}
\includegraphics[width=1.0\textwidth]{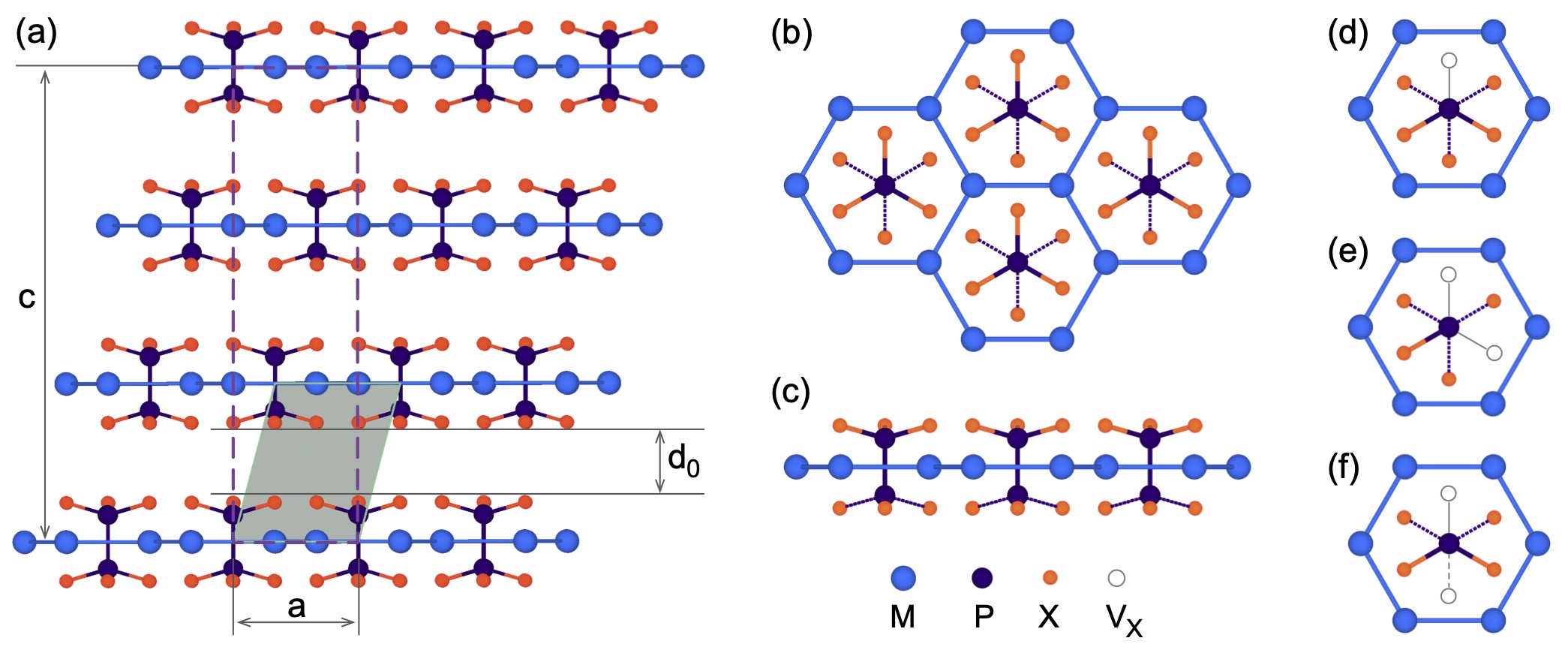}
\end{center}
\caption{(Left panel) (a) Crystal structure of 3D MPX$_3$. Dashed line denotes the hexagonal unit cell used in the present work. The in-plain and out-of-plane lattice constants are indicated with letters $a$ and $c$, respectively. Shaded area denotes a primitive unit cell.  Distance $d_0$ corresponds to the van-der-Waals gap. Spheres of different size/colour represent ions of different type. (Middle panel) Top (b) and side (c) views of a single layer of MPX$_3$. (Right panel) The considered chalcogen defects: (d) V$_\mathrm{X}$@1L; (e) V$_\mathrm{X2}$@1L; (f) V$_\mathrm{X2}$@2L. }
\label{fig:structure}
\end{figure}

Mechanical exfoliation used for monolayers production, can lead to formation of chalcogen vacancies. Therefore, we consider three different kinds of defects: (i) one vacancy at X-site, named as V$_\mathrm{X}$@1L with a defect concentration about $2.5$\,\% (Fig.~\ref{fig:structure}\,c); (ii) two vacancies at the neighbouring X-sites of the same chalcogen sublayer named as V$_\mathrm{X2}$@1L with a concentration about $5$\,\% (Fig.~\ref{fig:structure}\,d); (iii) two vacancies at X-sites of the different chalcogen sub-layers named as V$_\mathrm{X2}$@2L with a concentration about $5$\,\% (Fig.~\ref{fig:structure}\,e). The defect formation energies (Tab.~\ref{tab:defects}) are of the same order that is known for the other MPX$_3$ family members~\cite{Yang:2020ex,Wu:2021ac} and are rather low in all cases presumably indicating a feasibility of formation of these kinds of defects in experiments. According to the defect formation energies, it is found that V$_\mathrm{X}$@1L defects are the most favourable to form, followed by V$_\mathrm{2X}$@2L and V$_\mathrm{2X}$@1L. Also, we can observed that for the same defect type, Se vacancy is more likely to occur than S vacancy, which correlates with the respective electronegativity values. Although the rather low $\Delta E_\mathrm{def}$ obtained for all considered cases, we expect that in experimental conditions the formation of single chalcogen defects is more feasible and from now on we will focus on detailed consideration of V$_\mathrm{X}$@1L.

Removal of chalcogen atom leads to some modifications of the local lattice and electronic structures. As compared to the pristine CrPX$_3$ monolayer, the phosphorous dimers tend to moving closer to the vacancy. The angle, which P--P dimer forms with the vertical direction is $2.5^\circ$ and $3.9^\circ$ for X = S and Se, respectively. The dimer is pulled out from a monolayer. This leads to somewhat increase of the P--S bond length (by $\approx0.05$\,\AA), while the P--P bond lengths remain basically unchanged. 

\begin{figure}[h]
\begin{center}
\includegraphics[width=0.5\textwidth]{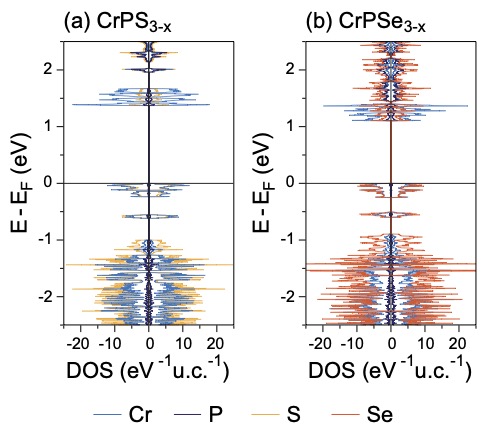}
\end{center}
\caption{Site-projected density of states for the defective monolayers: (a) 2D CrPS$_{3-x}$ and (b) 2D CrPSe$_{3-x}$.}
\label{fig:dos2D}
\end{figure}

The electrons left behind upon removal of a chalcogen atom, occupy the easily available electronic states of a (P$_2$X$_5$) entity. The magnetic moments of the Cr$^{2+}$-ions nearby the vacancy are coupled antiferromagnetically.  Similarly to the data recently published for MnPX$_3$~\cite{Yang:2020ex}, there are no well-localised defect state generated within the band gap in the DOS (Fig.~\ref{fig:dos2D}). As a result,  the energy gap width of the defective monolayers stays almost unchanged regarding the pristine CrPX$_3$: $E_g = 1.41$\,eV and $1.02$\,eV for the defective CrPS$_3$ and CrPSe$_3$, respectively. At the same time, at $E-E_F\approx-0.55$\,eV a new state appears, which is formed by S/Se $p$ and equal to that by the Cr $3d$ orbitals.

The estimated $T_\mathrm{N}$ value for the defective trichalcogenide monolayers is reduced to $158$\,K and $69$\,K for CrPS$_3$ and CrPSe$_3$, respectively (Fig.~\ref{fig:cleav+MC}). This effect is generally expected because  the X-vacancy formation leads to the drastic reduction of the $J$ exchange-coupling parameters ($J_1=-2.76$\,meV, $J_2=0.03$\,meV, $J_3=0.20$\,meV for CrPS$_{3-x}$; $J_1=-1.59$\,meV, $J_2=0.02$\,meV, $J_3=0.28$\,meV for CrPSe$_{3-x}$). At the same time the strongly localised magnetic moments of Cr$^{2+}$ ions remain the same. 

\subsection{Adsorption of H$_2$O }

The effect of adsorption of molecular water plays an important role in various applications and technological processes, since they are present in the environment of any device. Thus in the next step we investigate how adsorption of H$_2$O affects the electronic properties of CrPX$_3$ monolayers.

Various high-symmetry adsorption sites and adsorption orientations are taken into account. Considering the pristine monolayers, all adsorption configurations have similar adsorption energies which range from $-133$\,meV to $-153$\,meV and from $-148$\,meV to $-166$\,meV for 2D CrPS$_3$ and 2D CrPSe$_3$, respectively. In the relaxed structures water stays almost parallel to the substrate and the vertical distance between the molecule and the CrPX$_3$ monolayer is always above $3.2$\,\AA. 

\begin{figure}[h]
\includegraphics[width=1.0\textwidth]{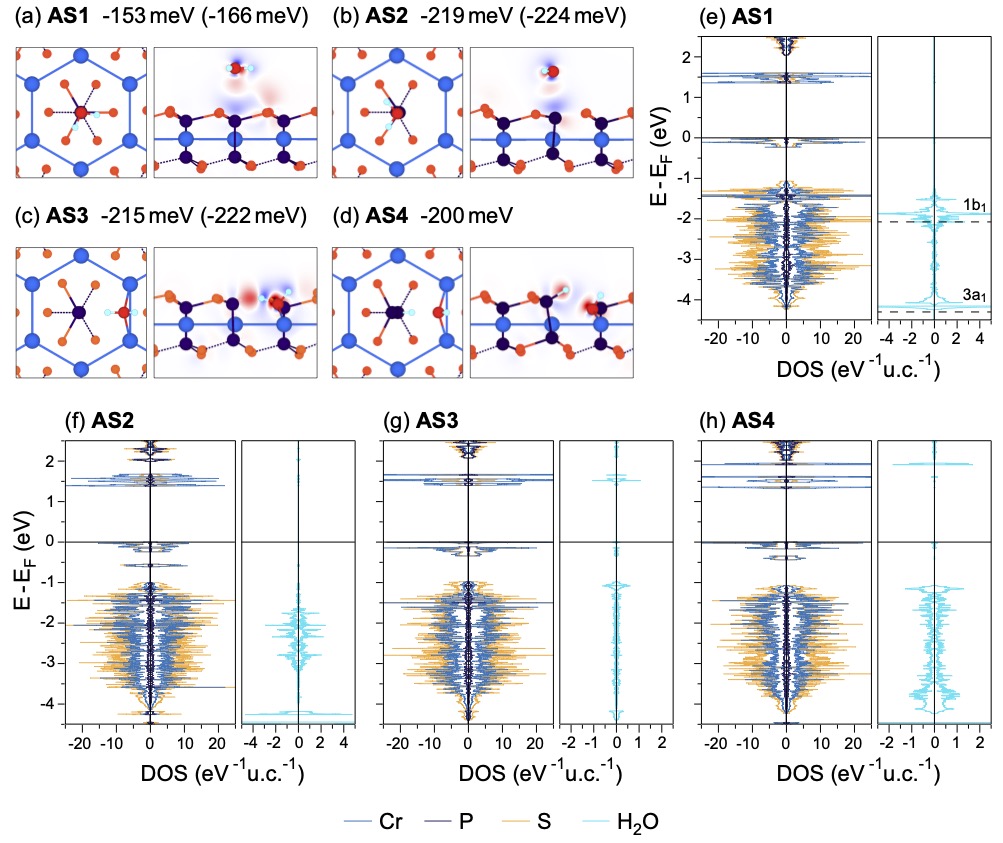}
\caption{(a-d) Top and side views of the relaxed structures obtained after water adsorption on pristine and defective CrPX$_3$ (cf. text for details). Side view are superimposed with electron density redistribution maps. Electron density accumulation (depletion) is shown in red (blue). Blue, violet, and orange spheres represent Cr, P, and X atoms, respectively. The water molecule is shown with red and light-blue spheres, for O and H, respectively. The adsorption energies are given above each adsorption structure for X=S (X=Se). (e-h) Site-projected density of states obtained for the structures presented in (a-d) where X = S. In (e), the molecular orbitals of a gas-phase H$_2$O molecule are indicated by horizontal dashed lines. These are obtained by aligning the $2a_1$ core level of the gas-phase and adsorbed H$_2$O molecules.}
\label{fig:H2O}
\end{figure}

For pristine monolayers, in the most stable adsorption structure (\textbf{AS1}, Fig.~\ref{fig:H2O}a), a water molecule attaches with its oxygen atom to an P atom [$d(\textrm{P-O}) = 3.23$\,\AA\ and $3.32$\,\AA\ for CrPS$_3$ and CrPSe$_3$, respectively] and the H atoms are directed towards the neighbouring X atoms. In accordance with the weak interaction, the structural parameters of H$_2$O as well as of the studied monolayer undergo insignificant changes. As follows from the electron density redistribution plot (Fig.~\ref{fig:H2O}a), upon adsorption the main charge rearrangement takes place between O and P atoms. Charge accumulation in the $p$-orbital of P on the side of the adsorbate indicates that it is the main orbital participating in the bonding. This charge accumulation is accompanied with a depletion at the hydrogen positions. These observation are in line with the effects arising in DOS due to the interaction between H$_2$O and CrPX$_3$ (Fig.~\ref{fig:H2O}e): Hybridisation between $p$ states of P and the water lone pair [$3a_1$ molecular orbital (MO) of H$_2$O] takes place, that results in broadening of the respective water derived states. Besides, a slight upward shift with respect to the MO of gas phase water molecule is observed. The obtained results on the adsorption of H$_2$O on pristine CrPX$_3$ are in general trend observed for graphene and for the other 2D chalcogenides. The adsorption energies for H$_2$O on graphene, MoS$_2$, and WS$_2$ are $-0.12...-0.14$\,eV, $-0.15$\,eV, and $-0.15$\,eV, respectively~\cite{Voloshina:2011ck,Levita:2016aaa,Bui:2015aaa}, indicating the physisorption nature of interaction. 

Considering the defective monolayers, the situation is more interesting. Firstly, the adsorption structure similar to \textbf{AS1} exists. It is abbreviated as \textbf{AS2} and presented in Figure~\ref{fig:H2O}b. In this case, the interaction between H$_2$O and the substate is stronger ($E_\mathrm{ads}=-219$\,meV and $E_\mathrm{ads}=-224$\,meV), which is reflected by the reduced distance $d(\textrm{P-O}) = 2.85$\,\AA\ and $2.98$\,\AA\ for CrPS$_3$ and CrPSe$_3$, respectively. In the rest, the electronic structures of the molecule and substate undergo qualitatively similar changes as in the case of \textbf{AS1}  (Fig.~\ref{fig:H2O}f). 

In addition to \textbf{AS2}, a structure, called \textbf{AS3} was investigated, where water molecule is coordinated between two Cr ions as it is shown in Figure~\ref{fig:H2O}c. (Similar configuration was found to be the energetically most favourable when studying water adsorption on  NiPX$_{3-x}$~\cite{Wu:2021ac}). Charge accumulation is observed between O and two Cr atoms as well as between H and P atom, which lost the chalcogen neighbour (Fig.~\ref{fig:H2O}c). Significant contribution of the the $1b_1$ and $3a_1$ MOs to the bonding with the monolayer is expressed in their substantial broadening due to hybridisation with P-$p$ sates. The healing of vacancy is accompanied by shift of the the defect derived state from its original position to lower binding energies where it is mixed with the valence band states. 

The interaction between between H and P atoms in \textbf{AS3} has one important effects on the geometric structure: A significant elongation of the H--O bond from $0.97$\,\AA\ for the gas-phase molecule to $1.06$\,\AA\ and $1.04$\,\AA\ for the adsorbed molecule on CrPS$_{3-x}$ and CrPSe$_{3-x}$, respectively. A further H--O bond elongation up to $d(\textrm{O-H}) = 2.02$\,\AA\ leads to structure \textbf{AS4} (Fig.~\ref{fig:H2O}d), which is slightly higher in energy as compared to \textbf{AS3}. Here the P--H bond is formed ($d(\textrm{P-H}) = 1.40$\,\AA). Similarly to recently published results for NiPX$_3$~\cite{Wu:2021ac}, the corresponding structure was found to be unstable for CrPSe$_{3-x}$. The weaker adsorbate-substrate interaction in the case of \textbf{AS4} is reflected by less pronounced changes in the DOS of the interacting species. By that the qualitative tendency of the induced modifications in \textbf{AS4} is similar to \textbf{AS3}.

\section{Conclusions}

In conclusions, we performed systematic DFT studies of pristine and chalcogen-defective 2D CrPX$_3$ (X: S, Se). It is shown that both pristine materials are wide gap semiconductors in the ground state with a band gap of $1.52$\,eV and $1.11$\,eV, respectively. The chalcogen vacancies are characterised by very strongly localised defect states located at $\approx0.5$\,eV below the valence band maximum and in the gap formed by valence band states, indicating the minimal influence of such defects on the optical and transport properties of CrPX$_3$. The adsorption of H$_2$O on the pristine and chalcogen-defective 2D CrPX$_3$ layers is always described as a physisorption, with possible water splitting. However, the clear preferential adsorption of H$_2$O on all kinds of surfaces of CrPX$_3$, molecular or dissociative, is not observed. It is found that in all cases of the surface modifications (chalcogen vacancies or ambient molecules adsorption), the electronic structure of CrPX$_3$ remains robust against these factors, indicating the possible application of these materials as effective protecting coatings.
 


\bigskip
\bigskip

\bibliographystyle{iopart-num.bst}

\providecommand{\newblock}{}

\clearpage
\begin{table}[h]
\caption{\label{tab:3D} Results obtained for the different magnetic states of 3D CrPS$_3$ and CrPSe$_3$ with PBE$+U$+D2: $E$ (in eV) is the total energy per unit cell, $a,c$ (in \AA) are the optimised lattice parameters, $d_0$  (in \AA)  is the van der Waals gap of bulk crystal. }
\begin{tabular*}{\textwidth}{@{}l*{15}{@{\extracolsep{0pt plus
12pt}}l}}
\br
X	&Symmetry	&Magnetic state	& $E$ 		&$a$		&$c$	 & 	$d_0$		\\		
\mr
S	&$C2/m$		&NM		&$-150.706$	&$5.75$	&$19.35$&$3.32$		\\	
	&			&FM		&$-167.336$	&$6.01$	&$19.48$&$3.27$		\\
	&			&nAFM	&$-168.501$	&$6.01$	&$19.48$&$3.21$		\\[0.3cm]
	&$R\bar3$	&NM		&$-150.083$	&$5.75$	&$19.35$&$3.33$		\\
	&			&FM		&$-167.300$	&$6.01$	&$19.48$&$3.26$	\\
	&			&nAFM	&$-166.507$	&$6.01$	&$19.48$&$3.21$		\\[0.3cm]		
Se	&$C2/m$		&NM		&$-134.846$	&$6.16$	&$19.44$&$3.30$	\\	
	&			&FM		&$-153.804$	&$6.35$	&$19.94$&$3.26$		\\
	&			&nAFM	&$-154.568$	&$6.35$	&$19.94$&$3.20$		\\[0.3cm]
	&$R\bar3$	&NM		&$-134.901$	&$6.16$	&$19.44$&$3.32$	\\
	&			&FM		&$-153.137$	&$6.35$	&$19.94$&$3.25$	\\
	&			&nAFM	&$-153.141$	&$6.35$	&$19.94$&$3.20$	\\[0.3cm]			
\br
\end{tabular*}
\end{table}

\clearpage
\begin{table}[h]
\caption{\label{tab:2D} Results obtained for the different magnetic states of 2D CrPS$_3$ and CrPSe$_3$ with PBE$+U$+D2:  $E_\mathrm{tot}$ (in eV per ($2\times 2$) unit cell) is the total energy; $\Delta E$ (in meV) is the difference between the total energy calculated for the magnetic states and the total energy calculated for the nAFM state. Band gap ($E_\mathrm{g}$, in meV), Cr$^{2+}$ magnetic moment ($M$, in $\mu_\mathrm{B}$),  the exchange coupling parameters between two local spins ($J$, in meV), and N\'eel temperature ($T_\mathrm{N}$, in K) are given for the lowest-energy structure. }
\begin{tabular*}{\textwidth}{@{}l*{15}{@{\extracolsep{0pt plus
12pt}}l}}
\br
X	&Magn. state	&$E_\mathrm{tot}$	&$\Delta E$	&$E_\mathrm{g}$	&$M$	&$J_1$	&$J_2$	&$J_3$	&$T_\mathrm{N}$		\\		
\mr
S	&FM			&$-223.260$		&$480$		&		&				&		&		&		&		\\
	&nAFM		&$-223.740$		&$0$			&$1.52$	&$3.8$			&$-3.10$	&$0.14$	&$0.32$	&$190$	\\
	&sAFM		&$-223.587$		&$153$		&		&				&		&		&		&	\\
	&zAFM		&$-223.351$		&$389$		&		&				&		&		&		&	\\[0.3cm]
Se	&FM			&$-204.172$		&$294$		&		&				&		&		&		&	\\
	&nAFM		&$-204.466$		&$0$			&$1.11$	&$3.8$			&$-2.06$	&$0.15$	&$0.36$	&$119$	\\
	&sAFM		&$-204.375$		&$91$		&		&				&		&		&		&	\\
	&zAFM		&$-204.194$		&$272$		&		&				&		&		&		&	\\[0.3cm]	
\br
\end{tabular*}
\end{table}

\clearpage
\begin{table}[h]
\caption{\label{tab:defects} Defect formation energies ($\Delta E_\mathrm{def}$, in eV)  obtained for the CrPX$_{3-x}$ monolayers. }
\centering
\begin{tabular*}{0.5\textwidth}{@{}l*{15}{@{\extracolsep{0pt plus
12pt}}c}}
\br
Defect type				&X = S &X = Se	 \\
\mr
V$_\mathrm{X}$@1L		&$1.069$	& $0.810$ \\
V$_\mathrm{X2}$@1L		&$1.222$	& $1.120$ \\
V$_\mathrm{X2}$@2L		&$1.098$	& $0.863$ \\
\br
\end{tabular*}
\end{table}

\clearpage

\clearpage

\clearpage

\clearpage

\clearpage

\clearpage

\end{document}